\documentclass[showpacs,prb,twocolumn,floats,superscriptaddress]{revtex4-2}
\usepackage{color}
\usepackage{amsmath,amssymb}
\usepackage{pifont}% http://ctan.org/pkg/pifont
\usepackage{amssymb}  % More symbols
\usepackage{bbold}
\usepackage{float}
\usepackage{tikz}
\usepackage{makecell}
\usepackage{subfigure}
\usepackage{pifont}   % Ding symbols
\usepackage{graphicx} % Include figure files
\graphicspath{{Figures/}}
\usepackage{dcolumn}  % Align table columns on decimal point
\usepackage{bm}       % bold math
\usepackage{multirow} % Table functions
\usepackage{placeins}% For making floats not move around everywhere
\usepackage[colorlinks]{hyperref}

\begin{document}

\title{Transport Properties in Gapped Bilayer Graphene }
\date{\today}
\author{N. Benlakhouy}
\email{benlakhouy.n@ucd.ac.ma}
\affiliation{Laboratory of Theoretical Physics, Faculty of Sciences, Choua\"ib Doukkali University, PO Box 20, 24000 El Jadida, Morocco}
\author{A. El Mouhafid}
\email{elmouhafid@gmail.com}
\affiliation{Laboratory of Theoretical Physics, Faculty of Sciences, Choua\"ib Doukkali University, PO Box 20, 24000 El Jadida, Morocco}
\author{A. Jellal}
\email{a.jellal@ucd.ac.ma}
\affiliation{Laboratory of Theoretical Physics, Faculty of Sciences, Choua\"ib Doukkali University, PO Box 20, 24000 El Jadida, Morocco}
%\affiliation{Saudi Center for Theoretical Physics, Dhahran, Saudi
%Arabia}
\affiliation{Canadian Quantum Research Center, 204-3002 32 Ave Vernon, \\ BC V1T 2L7, Canada}
\pacs{ 73.22.Pr, 72.80.Vp, 73.63.-b
}

\begin{abstract}

We  investigate  transport properties through a rectangular potential barrier in AB-stacked bilayer graphene (AB-BLG) gapped by dielectric layers. Using the Dirac-like Hamiltonian with a transfer matrix approach we obtain transmission and reflection probabilities as well as the associated  conductance. For  two-band model 
and at normal incidence, we find   extra resonances appearing in transmission   compared to biased AB-BLG, which are  Fabry-P\'erot resonance type. Now by taking into account the inter-layer bias, we show that both of transmission and 
 anti-Klein tunneling are diminished. Regarding   four band model, we find that
 the gap suppresses transmission in an energy range by showing some behaviors look like "Mexican hats". 
 We examine the total conductance and show that it is affected by the gap  compared to AA-stacked bilayer graphene. In addition, we find that the suppression in conductance is more important than that for biased AB-BLG.
\end{abstract}

\maketitle
\section{Introduction}
The experimental realization of monolayer graphene (MLG) in 2004  by Novoselov and 
Geim \cite{Novoselov} 
 opened up a
 new field in  physics. Such material has attractive electronic, optical, thermal, and mechanical properties. In particular, the observation of Klein tunneling \cite{Katsnelson,Klein}, anomalous quantum Hall effect \cite{Novoselov, Zhang}, and optical transparency \cite{Nair}. This makes graphene a good platform for nanoscale adaptor applications \cite{Neto}. Bilayer graphene (BLG) is a system formed by two stacked sheets of graphene. Besides that, there are two distinct kinds of stacking: AB-BLG or AB-(Bernal) \cite{Bernal}, and AA-BLG. AB-BLG has a parabolic dispersion relation with four bands where two of them touch at zero energy, whereas the other two bands  split together by the  interlayer hopping parameter $\gamma_{1}\approx0.4$ eV \cite{Li}. This structure is much more stable and its high-quality samples are developed and studied  theoretically and experimentally \cite{McCann, Rozhkov, Ohta, Goerbig, Jellal2}. AA-BLG has a linear energy gapless spectrum with two Dirac cones switched in energy by the quantity $\gamma_{1}\approx 0.2$ eV \cite{Labato}, and because of this AA-BLG attained enormous theoretical interest \cite{Rakhmanov, Mohammadi, Chen, Chiu, Redouani, Zahidi}. Such a structure is expected to be metastable, just lately, stable samples were discovered \cite{Lee, Borysiuk, Andres, Liu}. The AB-BLG may have clearly defined benefits than MLG, due to greater possibilities for balancing their physical properties. For reference: quantum Hall effect \cite{McCann, McCann1}, spin-orbit coupling and  transverse electric field \cite{Konschuh}, transmission probability in presence of electric and magnetic static fields \cite{Jellal1,Jellal2}, and quantum dots \cite{Giavaras}.

Experimentally, the evidence of  Klein tunneling in MLG was confirmed \cite{Ben,  Katsnelson,  Young,  Stander}, which means that there is no electron confinement, and then a gap must be created to overcome this issue. In fact,  many methods of induction a band gap in MLG have been elaborated such as substrates \cite{Wang, Jose, Kindermann, Song, Jung, Nevius, Zarenia, Uchoa} and doping with impurities \cite{Zhou, Costa}. Regarding AB-BLG, band gap can be realized by  applying an external electric field \cite{Tang, McCann} or induced by using dielectric materials like hexagonal boron nitride (h-BN) or SiC \cite{Zhai}. %\cite{Bahlouli}. 
To this end, it is theoretically showed that %the band gap can If we consider the mass term in both layers then 
 quantum spin Hall phase can be identified in gapped AB-BLG even when the Rashba interaction approached zero \cite{Zhai}.

The introduction of an inter-layer bias to  AB-BLG  opens a gap in the energy spectrum and has a major effect on electronic properties \cite{Ben}. Here, we analyze the effects of a biased AB-BLG gapped by dielectric layers   to show the impact of  band gap on  transport  properties. In both layers of AB-BLG,  band gap is the same
allowing   to open a gap. 
 Using transfer matrix method together with current density, we calculate  transmission and reflection probabilities as well as  corresponding conductance. 
%{\rr At low-energy, $E<\gamma_{1}$, and in presence of the band gap $\Delta_{0}$   we find that %show that when we introduce a band gap (denoted by $\Delta_{0}$) 
%	Fabry-Pérot resonances \cite{Snyman} strongly appear in the transmission. Now by including also the inter-layer bias $\delta$, we show that  total transmission 
%	and anti-Klein tunneling significantly diminished. For energies exceeding the inter-layer coupling $\gamma_{1}$,  $E>\gamma_{1}$, we obtain two propagating modes giving rise to the four transmission channels. 
	%Under suitable conditions of physical parameters, %We are illustrating that the application of the band gap opens a gap and 
%In the energy range $V_0-\Delta_{0}<E<V_0+\Delta_{0}$, we notice that presence of $\Delta_{0}$	suppresses the transmission and breaks the momentum  symmetry. Additionally, we present different numerical results and
%compare them with published works on the topic. In particular, we observe extra resonances in transmissions and conductance got modified due to band gap.}
At low-energy, $E<\gamma_{1}$, and in presence of the band gap $\Delta_{0}$ we find that 
	Fabry-Pérot resonances \cite{Snyman} strongly appear in the transmission. Now by including also the inter-layer bias $\delta$, we show that the total transmission and anti-Klein tunneling significantly diminished. For energies exceeding the inter-layer coupling $\gamma_{1}$,  $E>\gamma_{1}$, we obtain  a new mode of propagating  giving rise to the four transmission channels. In this case,  $\Delta_{0}$ suppresses the transmission
	in the energy range $V_0-(\Delta_{0}+\delta)<E<V_0+(\Delta_{0}+\delta)$, 
	and shows some behaviors that look like ``Mexican hats". 
	%We found that the band gap $\Delta_{0}$  breaks the momentum  symmetry. 
	Finally we find that the resulting conductance in  gapped AB-BLG gets modified compared to gapped AA-BLG. Moreover, we find that the suppression in conductance is more important than that  for  biased AB-BLG  \cite{Ben} because the energy range for a null conductance increases as long as $\Delta_{0}$ increase and also the number of peaks get reduced. 

The paper is organized as follows. In Sec \ref{Theoretical model} we  construct our theoretical model describing biased and gapped AB-BLG giving rise to four band energies. In Sec \ref{Transmission probability and conductance} we explain in detail the formalism used
in calculating  transmission and reflection probabilities together with conductance. In Sec \ref{NUMERICAL RESULTS AND DISCUSSION} we  numerically analyze our  results and give different discussions  with published works on the topic. Finally, in Sec. \ref{Summary and conclusion} we summarize our main conclusions.
%%%%%%%%%%%%%%%%%%%%%%%%%%%%%%%%%%%%%%
\section{Theoretical model}
%%%%%%%%%%%%%%%%%%%%%%%%%%%%%%%%%
%
\label{Theoretical model}
\begin{figure}[h!] \centering
\includegraphics[width=3.2 in]{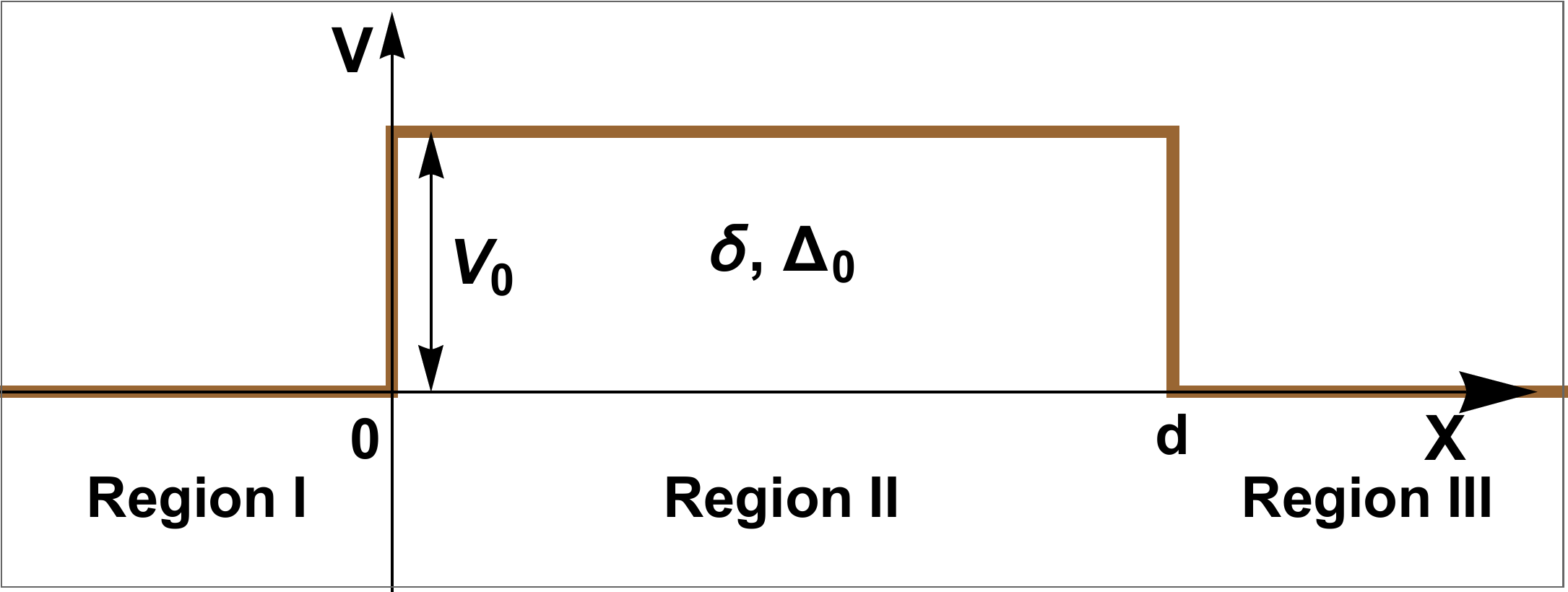} \caption{The parameters
of a rectangular barrier structure.}\label{barrier}
\end{figure}
In the AB-stacked bilayer graphene % there is two single layers of graphene which 
the atom $B_{1}$ of the top layer is placed directly  below the atom $A_{2}$ of  the bottom layer  with van der Waals inter-layer coupling parameter $\gamma_{1}$,  while $A_{1}$ and $B_{2}$ do not lie directly below or above each other. Based on \cite{Ben,Zhai} we consider a biased and gapped AB-BLG 
%graphene  in the presence of a mass term, 
described by the following   Hamiltonian near the point \textbf{K} %as follows
\begin{equation}\label{1}
\mathcal{H}=\begin{pmatrix}
V_0+\vartheta_1 & v_{F}\pi^{\dag} & -v_{4}\pi^{\dag} & v_{3}\pi \\
v_{F}\pi &  V_0+\vartheta_2 & \gamma_{1} & -v_{4}\pi^{\dag}\\
-v_{4}\pi & \gamma_{1} & V_0-\vartheta_2 & v_{F}\pi^{\dag} \\
v_{3}\pi^{\dag} & -v_{4}\pi &  v_{F}\pi& V_0-\vartheta_1 \\
\end{pmatrix}
\end{equation}
where $ v_{F}=\frac{\gamma_{0}}{\hbar} \frac{3 a}{2}\approx10^{6}$\
{m}/{s} is the Fermi velocity of electrons in each graphene
layer, $a=0.142$\ nm is the distance between adjacent carbon
atoms, $v_{3,4}=\frac{v_{F} \gamma_{3,4}}{\gamma_{0}}$ represent
the coupling between the layers, $\pi=p_{x}+ip_{y},
\pi^{\dag}=p_{x}-ip_{y}$ are the in-plan momenta and its conjugate
with $p_{x,y}=-i\hbar
\partial_{x,y}$, %+\frac{e}{c}A$ and $A$ is the vector potential.
$\gamma_{1}\approx 0.4$\ eV is the interlayer coupling term.   The electrostatic potential $V_0$ of width $d$  (Fig. \ref{barrier}) %, on the i-th layer which 
can be varied on the $i$-th layer using top and back gates on the sample. $\vartheta_1=\delta+\Delta_{0}$, $\vartheta_2=\delta-\Delta_{0}$ with $\delta$ corresponds to an externally induced inter-layer potential difference, and $\Delta_{0}$ is the band gap. The skew parameters, $\gamma_3\approx0.315$\ eV and
$\gamma_4\approx0.044$\ eV have negligible effect on the band
structure at high energy \cite{McCann1,McCann2}. Recently, it was shown that
even at low energy these parameters have also negligible effect on
the transmission \cite{Ben}, hence we neglect them in our
calculations. 

Under the above approximation and for a barrier  potential
configuration as depicted in Fig. \ref{barrier}, the Hamiltonian (\ref{1}) can be written as
\begin{equation}\label{eq2}
H=\left(
\begin{array}{cccc}
  V_0+\vartheta_1 & \nu_{F}\pi^{\dag} & 0 & 0 \\
  \nu_{F}\pi & V_0+\vartheta_2 & \gamma_{1} & 0\\
  0 & \gamma_{1} & V_0-\vartheta_2 & \nu_{F}\pi^{\dag} \\
  0 & 0& \nu_{F}\pi & V_0-\vartheta_1 \\
\end{array}%
\right)
\end{equation}

By considering the length scale $l=\hbar v_{F}/\gamma_1$,
which represents the inter-layer coupling length $l=1.64$ nm,
%allows us to 
we define the dimensionless quantities: 
$x\equiv x/l$ and $k_y\equiv l k_y$ together with 
$\delta\equiv\frac{\delta}{\gamma_1}$, $\Delta_{0}\equiv\frac{\Delta_{0}}{\gamma_1}$, $E\equiv\frac{E}{\gamma_1}$, $V_0\equiv\frac{V_0}{\gamma_1}$. % {\rr Note that in region I and III we have $V_0=\delta=\Delta_0=0$.}
The eigenstates of Eq. (\ref{eq2}) are four-components spinors
$\psi(x,y) =[{\psi}_{A_{1}},{\psi}_{B_{1}},{\psi}_{A_{2}},{\psi}_{B_{2}}]^{\dag}$,
here $\dag$ denotes the transpose of the row vector.
% and $j$ denotes the $j$-th region. 
As a consequence  of the transnational invariance along the $y$-direction, %the momentum along the $y$-direction is a conserved quantity, i.e $
we have $[H,p_{y}]=0$, and then we  decompose the spinor as
\begin{equation}\label{2}
\psi(x,y)=e^{ik_{y}y}\left[\phi_{A_{1}}(x),\phi_{B_{1}}(x),\phi_{A_{2}}(x),\phi_{B_{2}}(x)\right]^{T}
\end{equation}

We solve  the time-independent Schr\"odinger equation $H\psi=E\psi$ to obtain a general
solution in the region II and then require $V_0=\delta=\Delta_0=0$ to derive the solutions in the regions I and III. Indeed, by substituting Eq. (\ref{eq2}) and Eq. (\ref{2}) we get four related differential equations 
\begin{subequations}\label{sys1}
 \begin{eqnarray}
-i(\partial_{x}+k_{y})\phi_{B_{1}} &=&\varepsilon_1\phi_{A_{1}} \label{eqs1}  \\
-i(\partial_{x}-k_{y})\phi_{A_{1}} &=&\varepsilon_2\phi_{B_{1}}-\phi_{A_{2}} \label{eqs2}  \\
-i(\partial_{x}+k_{y})\phi_{B_{2}}&=&\varepsilon_3\phi_{A_{2}}-\phi_{B_{1}} \label{eqs3}  \\
-i(\partial_{x}-k_{y})\phi_{A_{2}}&=&\varepsilon_4\phi_{B_{2}} \label{eqs4}
\end{eqnarray}
\end{subequations} 
where we have set $\varepsilon_1=\varepsilon-\vartheta_1$, $\varepsilon_2=\varepsilon-\vartheta_2       $, $\varepsilon_3=\varepsilon+\vartheta_2$, $\varepsilon_4=\varepsilon+\vartheta_1$ and $\varepsilon=E-V_0$. We solve Eq. (\ref{eqs1}) for $\phi_{A_{1}}$, Eq. (\ref{eqs4}) for $\phi_{B_{2}}$ and substitute the results in Eqs. (\ref{eqs2},\ref{eqs3}). This process yields 
\begin{subequations}\label{sys2}
 \begin{eqnarray}
(\partial_{x}^2-k_{y}^2+\varepsilon_1\varepsilon_2)\phi_{B_{1}} &=&\varepsilon_1\phi_{A_{2}} \label{eqss1}  \\
(\partial_{x}^2-k_{y}^2+\varepsilon_3\varepsilon_4)\phi_{A_{2}} &=&\varepsilon_4\phi_{B_{1}} \label{eqss2}\end{eqnarray}
\end{subequations}
%For the system of Eqs.(\ref{sys2}) and 
Then for constant parameters,  the energy bands are solution of the following  equation
\begin{equation}\label{3}
\left[-k^2+\varepsilon_1\varepsilon_2\right]\left[-k^2+\varepsilon_3\varepsilon_4\right]-\varepsilon_1\varepsilon_4=0
\end{equation}
such that  $k=\sqrt{k_x^2+k_y^2}$ and
the four possible wave vectors are given by
\begin{equation}
k^{s}_{x}=\sqrt{-k_{y}^{2}+ \varepsilon^{2}+\delta^{2}-\Delta_{0}^{2}\pm\sqrt{\varepsilon^{2}(1+4\delta^{2})-(\delta+\Delta_0)^{2}}}
\end{equation}
where $s =\pm$ defines the modes of propagation, which will be discussed in numerical section. Therefore,  the four energy bands   can be derived as 
\begin{widetext}
\begin{eqnarray}
\varepsilon^{s}_\pm=s\sqrt{k^{2}+\delta^{2}+\Delta_{0}^{2}+\frac{1}{2}\pm\sqrt{k^{2}\left(1+4\delta^{2}\right)+\left(\frac{1}{2}-2\delta\Delta_{0}\right)^{2}}}\label{energy}
%\\ &&\varepsilon^-_\pm=-\sqrt{k^{2}+\delta^{2}+\Delta_{0}^{2}+\frac{1}{2}\pm\sqrt{k^{2}\left(1+4\delta^{2}\right)+\left(\frac{1}{2}-2\delta\Delta_{0}\right)^{2}}}
\end{eqnarray}
\end{widetext}
At this level, we have some comments in order. Indeed, firstly
by taking $\delta=0$, \eqref{energy} reduces
\begin{equation}
	\varepsilon^{s}_\pm|_{\delta=0}=s\sqrt{k^{2}+\Delta_{0}^{2}+\frac{1}{2}\pm\sqrt{k^{2}+\frac{1}{4}}}\label{energy0}
\end{equation}
Secondly for the case $\Delta_0=0$, we end up with Ben {\it et al.} result \cite{Ben}
\begin{eqnarray}
	\varepsilon^{s}_\pm|_{\Delta_0=0}=
	s\sqrt{k^{2}+\delta^{2}+\frac{1}{2}\pm\sqrt{k^{2}\left(1+4\delta^{2}\right)+\frac{1}{4}}}
	\label{energy1}
\end{eqnarray}
Now by comparing   \eqref{energy0} and  \eqref{energy1}, we  clearly notice that both quantities $\delta$ and $\Delta_0$ are inducing different gaps in the energy spectrum. Certainly this
difference will affect the transmission probabilities (Figs. \ref{fig0203}, \ref{fig0204}) as well as conductance (Fig. \ref{fig6}).

\begin{figure}[tp]
\begin{center}
\includegraphics[width=6cm]{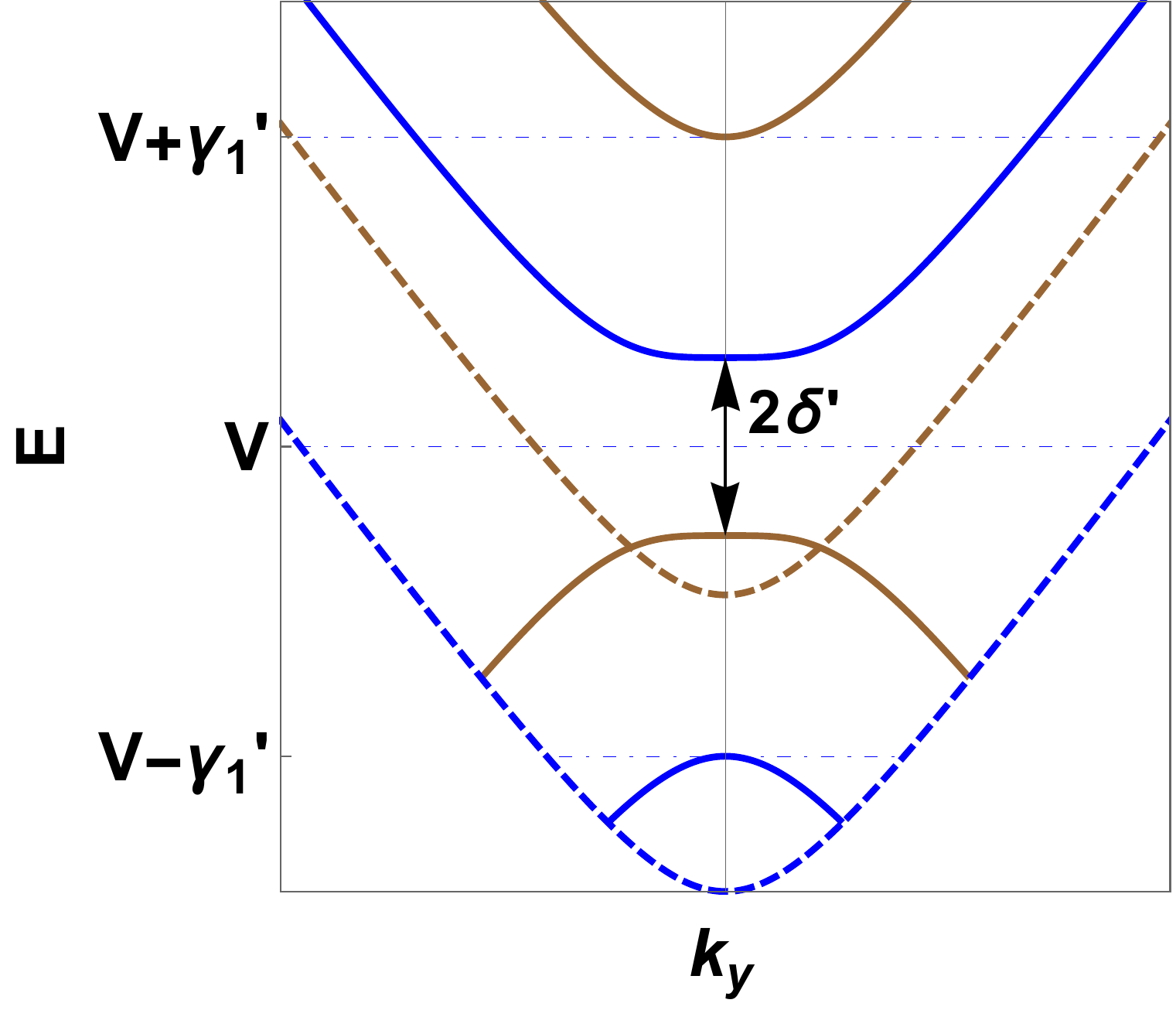}
\end{center}
\caption{Energy spectrum of AB-stacked graphene bilayer inside (solid curves) and outside (dashed curves) the barrier. Here blue
(brown) curves correspond to $k^+(k^-)$ propagating modes for biased and gapped
$(V_0\neq0, \delta\neq0, \Delta_{0}\neq0)$ systems. $\delta'=\delta+\Delta_{0}$ and $\gamma_1'=\sqrt{\gamma_1^2+(\delta-\Delta_{0})^2}$.}\label{fig0103}
\end{figure}

It is known that the perfect AB-BLG has a parabolic dispersion relation with four bands, of which two touch each other at $k=0$. In Fig. \ref{fig0103} we show the energy bands as a function of the momentum $k_{y}$, for the biased and gapped AB-BLG. We observe that when the AB-BLG is subjected to a gap $\Delta_{0}$ and an inter-layer bias $\delta$ the two bands are switched and placed at $V_{0}\pm\sqrt{\gamma_{1}^{2}+(\delta-\Delta_{0})^2}$, and the touching bands are shifted by $2\delta'=2(\delta+\Delta_{0})$. One should notice that there are two cases related to whether  the wave vector  $k^{s}_{0}=\sqrt{-k_{y}^{2}+\varepsilon^{2}\pm\varepsilon}$ is real or imaginary. Indeed for $E<\gamma_{1}$, just $k^{+}_{0}$ is real, and for that reason, the propagation is only possible for  $k^{+}_{0}$ mode. However when  $E>\gamma_{1}$, both  $k^{\pm}_{0}$ are real which presenting a new propagation mode.

As concerning the eigenspinors in regions II, we show that the solution of Eqs. (\ref{sys2})  is a plane wave generated by
\begin{equation}
\phi_{B_{1}}^2=a_{1}e^{ik_x^{+}x}+a_{2}e^{-ik_x^{+}x}+a_{3}e^{ik_x^{-}x}+a_{4}e^{-ik_x^{-}x}
\end{equation}
where $a_{n}$ are coefficients of normalization, with $ n = 1,\cdots, 4$. The remaining   components of the eigenspinors can be obtained as
\begin{widetext}
\begin{align}
\phi_{A_{1}}^2&=a_{1}\Lambda^{+}_{+}e^{ik_x^{+}x}+a_{2}\Lambda^{+}_{-}e^{-ik_x^{+}x}+a_{3}\Lambda^{-}_{+}e^{ik_x^{-}x}+a_{4}\Lambda^{-}_{-}e^{-ik_x^{-}x}\\
\phi_{A_{2}}^2&=a_{1}\rho^{+}e^{ik_x^{+}x}+a_{2}\rho^{+}e^{-ik_x^{+}x}+a_{3}\rho^{-}e^{ik_x^{-}x}+a_{4}\rho^{-}e^{-ik_x^{-}x}\\
\phi_{B_{2}}^2&=a_{1}\chi^{+}_{+}\rho^{+}e^{ik_x^{+}x}+a_{2}\chi^{+}_{-}\rho^{+}e^{-ik_x^{+}x}+a_{3}\chi^{-}_{+}\rho^{-}e^{ik_x^{-}x}+a_{4}\chi^{-}_{-}\rho^{-}e^{-ik_x^{-}x}
\end{align}
\end{widetext}
where we have introduced the quantities $\Lambda^{\pm}_{\pm}=\frac{-ik_{y}\pm k_x^{\pm}}{\varepsilon-\vartheta_1}$,  $\rho^{\pm}=\frac{(\epsilon-\vartheta_1)(\epsilon-\vartheta_2)-k_y^2-(k_x^{\pm})^2}{\epsilon-\vartheta_1}$, $\chi^{\pm}_{\pm}=\frac{ik_{y}\pm k_x^{\pm}}{\varepsilon+\vartheta_1}$.
In matrix notation, the general solution of our system in region II can be written 
 as
\begin{equation}
\psi_2(x,y)=\mathcal{G}_2\cdot\mathcal{M}_2(x)\cdot\mathcal{C}_2\ e^{ik_{y}y}
\end{equation}
where the four-component vector ${\cal{C}}_2$ represents the  coefficients $a_n$ expressing the relative weights of the different traveling modes, which have to be set according to the propagating region \cite{Ben}. The matrices $\mathcal{M}_2(x)$ and $\mathcal{G}_2$ are given by
\begin{widetext}
\begin{equation}
\mathcal{G}_2=\begin{pmatrix}
1 & 1 & 1 & 1\\
\Lambda^{+}_{-} & \Lambda^{+}_{+} & \Lambda^{-}_{+} & \Lambda^{-}_{-} \\
\rho^{+} & \rho^{+} & \rho^{-} & \rho^{-} \\
\chi^{+}_{+}\rho^{+} & \chi^{+}_{-}\rho^{+} &  \chi^{-}_{+}\rho^{-}& \chi^{-}_{-}\rho^{-} \\
\end{pmatrix},\qquad \mathcal{M}_2(x)=\begin{pmatrix}
e^{ik_x^{+}x} & 0 & 0& 0 \\
0 &e^{-ik_x^{+}x} & 0 & 0\\
0 &0  & e^{ik_x^{-}x} & 0\\
0 &0&  0& e^{-ik_x^{-}x} \\
\end{pmatrix},\qquad \mathcal{C}_2=\begin{pmatrix}
a_1 \\
a_2\\
a_3\\
a_4\\
\end{pmatrix}
\end{equation}
\end{widetext}

%The above solution coincides with intermediate region II, 
As claimed before, to get  solutions in the other regions we have to set  $V_0=\delta=\Delta_{0}=0$. Then the eigenspinors in region I is
\begin{widetext}
\begin{align}
% \nonumber % Remove numbering (before each equation)
\phi_{A_{1}}^{1}&=\delta_{s,1}e^{ik^{+}_{0}x}+r^{s}_{+}e^{-ik^{+}_{0}x}+\delta_{s,-1}e^{ik^{-}_{0}x}+r^{s}_{-}e^{-ik^{-}_{0}x}\\
\phi_{B_{1}}^{1}&=\delta_{s,1}\Lambda^{+}_{-}e^{ik^{+}_{0}x}+r^{s}_{+}\Lambda^{+}_{+}e^{-ik^{+}_{0}x}+\delta_{s,-1}\Lambda^{-}_{+}e^{ik^{-}_{0}x}+r^{s}_{-}\Lambda^{-}_{-}e^{-ik^{-}_{0}x}\\
\phi_{A_{2}}^{1}&=\delta_{s,1}\rho^{+}e^{ik^{+}_{0}x}+r^{s}_{+}\rho^{+}e^{-ik^{+}_{0}x}+\delta_{s,-1}\rho^{-}e^{ik^{-}_{0}x}+r^{s}_{-}\rho^{-}e^{-ik^{-}_{0}x}\\
\phi_{B_{2}}^{1}&=\delta_{s,1}\chi^{+}_{+}\rho^{+}e^{ik^{+}_{0}x}+r^{s}_{+}\rho^{+}\chi^{+}_{-}e^{-ik^{+}_{0}x}+\delta_{s,-1}\rho^{-}\chi^{-}_{+}e^{ik^{-}_{0}x}+r^{s}_{-}\rho^{-}\chi^{-}_{-}e^{-ik^{-}_{0}x}
\end{align}
and in the region III reads as 
\begin{align}
% \nonumber % Remove numbering (before each equation)
\phi_{A_{1}}^{3}&=t^{s}_{+}e^{ik^{+}_{0}x}+t^{s}_{-}e^{ik^{-}_{0}x}\\
\phi_{B_{1}}^{3}&=t^{s}_{+}\Lambda^{+}_{-}e^{ik^{+}_{0}x}+t^{s}_{-}\Lambda^{-}_{+}e^{ik^{-}_{0}x}\\
\phi_{A_{2}}^{3}&=t^{s}_{+}\rho^{+}e^{ik^{+}_{0}x}+t^{s}_{-}\rho^{-}e^{ik^{-}_{0}x}\\
\phi_{B_{2}}^{3}&=t^{s}_{+}\chi^{+}_{+}\rho^+ e^{ik^{+}_{0}x}+t^{s}_{-}\chi^{-}_{+}\rho_-e^{ik^{-}_{0}x}
\end{align}
\end{widetext}
Since the potential is zero in regions I and III, we have the
relation $\mathcal{G}_{1} \cdot\mathcal{M}_{1}(x)=\mathcal{G}_{3}\cdot \mathcal{M}_{3}(x)$. We will see how the above results will be used to determine different physical quantities. Specifically, we focus on the transmission and  reflection probabilities as well as the conductance.

\section{Transmission probability and conductance}
\label{Transmission probability and conductance}
To determine the transmission and  reflection  probabilities, we impose the appropriate boundary conditions in the context of the transfer matrix approach \cite{Barbier,Barbier1}. Continuity of the spinors at interfaces  gives the components of the vectors % $\mathcal{C}$ 
%which are generated by 
\begin{equation}
\mathcal{C}_{1}^{s}=\begin{pmatrix}
\delta_{s,1} \\
r_{+}^{s}\\
\delta_{s,-1}\\
r_{-}^{s} \\
\end{pmatrix},\qquad \mathcal{C}_{3}^{s}=\begin{pmatrix}
t_{+}^{s} \\
0\\
t_{-}^{s}\\
0 \\
\end{pmatrix}
\end{equation}
where $\delta_{s,\pm}$ is the Kronecker symbol. The continuity  at $x=0$ and $x=d$  can be written in a matrix notation as
\begin{eqnarray}
&& \mathcal{G}_{1}\cdot \mathcal{M}_{1}(0)\cdot \mathcal{C}_{1}^{s}=\mathcal{G}_{2} \cdot \mathcal{M}_{2}(0) \cdot \mathcal{C}_{2}\\
&&
\mathcal{G}_{2}\cdot \mathcal{M}_{2}(d) \cdot \mathcal{C}_{2}=\mathcal{G}_{3} \cdot \mathcal{M}_{3}(d) \cdot \mathcal{C}_{3}^{s}
\end{eqnarray}
Using the transfer matrix method together with the relation $\mathcal{G}_{1}\cdot \mathcal{M}_{1}(x)=\mathcal{G}_{3}\cdot \mathcal{M}_{3}(x)$ we can connect $\mathcal{C}_{1}^{s}$ with $\mathcal{C}_{3}^{s}$  through the matrix $\mathcal{N}$
\begin{equation}
\mathcal{C}_{1}^{s}=\mathcal{N} \cdot \mathcal{C}_{3}^{s}
\end{equation}
where
\begin{equation}
\mathcal{N}= \mathcal{G}_{1}^{-1}\cdot \mathcal{G}_{2} \cdot \mathcal{M}_{2}^{-1}(d)\cdot \mathcal{G}_{2}^{-1}\cdot \mathcal{G}_{1}\cdot \mathcal{M}_{1}(d)
\end{equation}
Consequently, the transmission and reflection coefficients can be derived from
\begin{equation}
\left(
\begin{array}{cccc}
t_{+}^{s} \\
r_{+}^{s}\\
t_{-}^{s}\\
r_{-}^{s} \\
\end{array} 
\right)=\left(
\begin{array}{cccc}
\mathcal{N}_{11} & 0 & \mathcal{N}_{13}& 0 \\
\mathcal{N}_{21} &-1 & \mathcal{N}_{23} & 0\\
\mathcal{N}_{31} &0  & \mathcal{N}_{33} & 0\\
\mathcal{N}_{41} &0&  \mathcal{N}_{43}& -1 \\
\end{array} 
\right)^{-1} \left(
\begin{array}{cccc}
 \delta_{s,1}\\
0\\
\delta_{s,-1}\\
0 \\
\end{array} 
\right)
\end{equation}
where $\mathcal{N}_{ij}$ are the matrix elements of  $\mathcal{N}$. Then, after some algebras,  we obtain the transmission and reflection coefficients %can be obtained as
\begin{align}
% \nonumber % Remove numbering (before each equation)
t_{+}^{s}&=\frac{\delta_{s,-1}\mathcal{N}_{13}-\delta_{s,1}\mathcal{N}_{33}}{\mathcal{N}_{13}\mathcal{N}_{31}-\mathcal{N}_{11}\mathcal{N}_{33}},\qquad t_{-}^{s}=\frac{-\delta_{s,-1}\mathcal{N}_{11}+\delta_{s,1}\mathcal{N}_{31}}{\mathcal{N}_{13}\mathcal{N}_{31}-\mathcal{N}_{11}\mathcal{N}_{33}}\\
% \nonumber % Remove numbering (before each equation)
r_{+}^{s}&=\mathcal{N}_{21}t_{+}^{s}+\mathcal{N}_{23}t_{-}^{s},\qquad r_{-}^{s}=\mathcal{N}_{41}t_{+}^{s}+\mathcal{N}_{43}t_{-}^{s} 
\end{align}

To calculate the transmission and reflection probabilities, we have to take  into account the change in velocity of the waves when they are scattered into a different propagation mode. For this, it is convenient to
use the current density $\boldsymbol{J}$ 
\begin{equation}
\boldsymbol{J}=v_{F}\boldsymbol{\psi}^{\dagger}\begin{pmatrix}
\sigma_{x} & 0  \\
0 & \sigma_{x} \\
\end{pmatrix}\boldsymbol{\psi}\label{6}
\end{equation}
where $\sigma_{x}$ is the Pauli matrix. Then Eq. (\ref{6}) gives
 the incident $\boldsymbol{J}_{x}^{\text{inc}}$, reflected $\boldsymbol{J}_{x}^{\text{ref}}$ and transmitted $\boldsymbol{J}_{x}^{\text{tra}}$ current densities. Finally the transmission $T$ and reflection $R$ probabilities are 
\begin{equation}
T^{s}_{\pm}=\frac{k^{\pm}_{0}}{k^{s}_{0}}|t^{s}_{\pm}|^{2}, \qquad R^{s}_{\pm}=\frac{k^{\pm}_{0}}{k^{s}_{0}}|r^{s}_{\pm}|^{2}
\end{equation}
To preserve the probability of current,  $T$ and  $R$  are normalized  as
 \begin{equation}
\sum_{i,j}\left(T^{j}_{i}+R^{j}_{i}\right)=1 
 \end{equation}
where the index $i=\pm$ points to the arriving mode, when the index $j=\pm$ points to the exiting mode. For example in the case of channel $k^{+}$, gives $T^{+}_{+}+T^{-}_{+}+R^{+}_{+}+R^{-}_{+}=1$. As already mentioned, for $E>\gamma_{1}$ we have two modes of propagation 
($k^{+}_{0}, k^{-}_{0}$) leading to four transmissions $T^{s}_{\pm}$ and four reflections $R^{s}_{\pm}$ channels, through the four conduction bands. For sufficiently enough low energy or in the two-band model, $E<\gamma_{1}$, the two modes lead to one transmission $T$ channel and one reflection $R$ channel.
 
 From the transmission probabilities, we can calculate the
 conductance $G$, at zero temperature, using  the Landauer-B\"{u}ttiker formula
 \begin{equation}
 G(E)=G_{0} \frac{L_{y}}{2 \pi} \int_{-\infty}^{\infty} d k_{y} \sum_{i, j=\pm} T_{i}^{j}\left(E, k_{y}\right)
 \end{equation}
 with $L_{y}$ the length of the sample in the $y$-direction, and $G_{0} = 4e^{2}/h$.  The factor $4$ comes from the valley and spin degeneracies in graphene. In order to get the total conductance of the system, we need to sum over all the transmission channels
 \begin{equation}
 G_{T}=\sum_{i,j}G^{j}_{i}
 \end{equation}
%%%%%%%%%%%%%%%%%%%%%%%%%%%%%%%%%%%%%%
\section{NUMERICAL RESULTS AND DISCUSSION}
\begin{figure}[tb]
\begin{center}
\includegraphics[width=9cm]{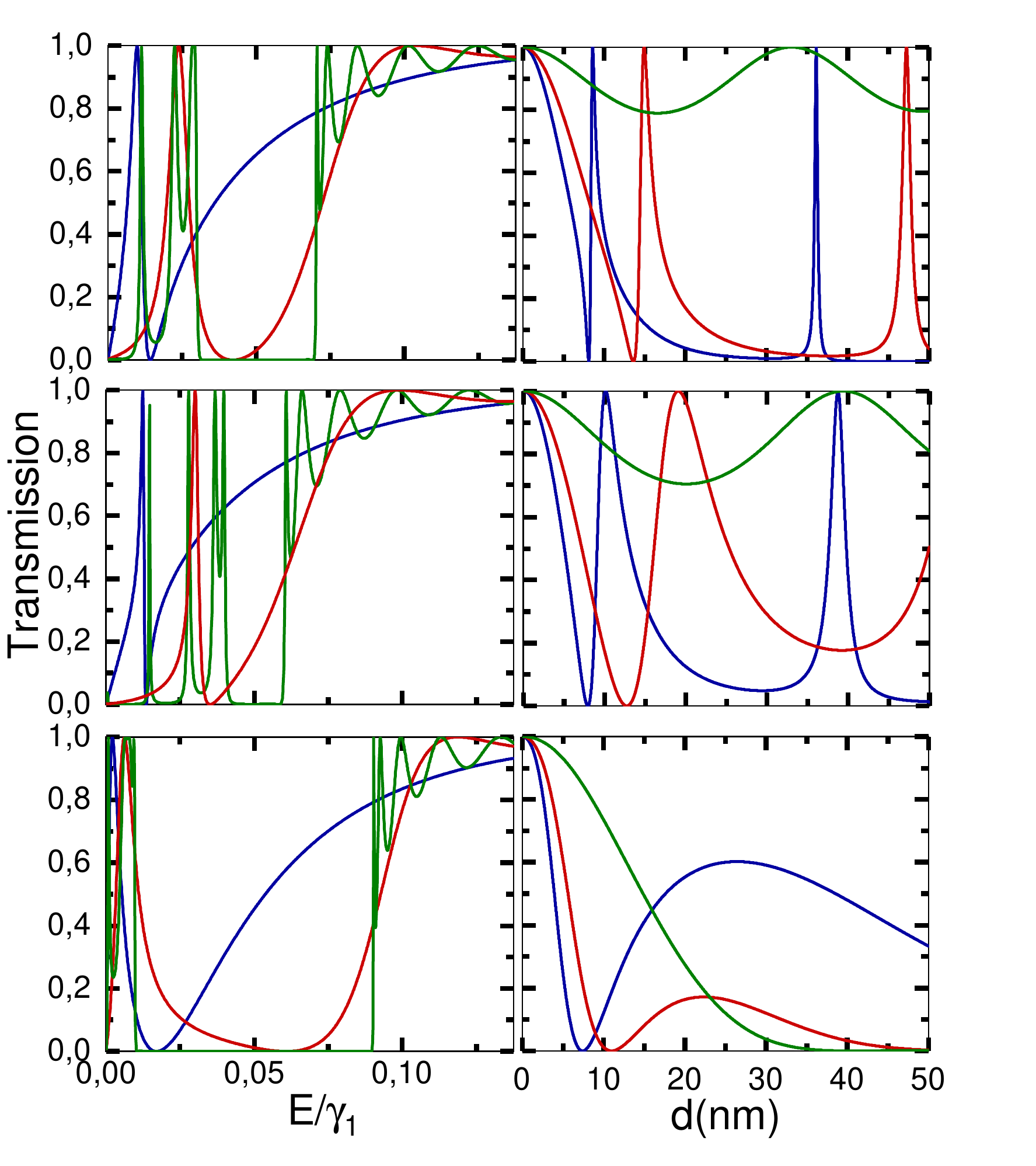}
\end{center}
\caption{(Color online) The transmission probability at normal incidence through a barrier of height $V_{0}=0.05\gamma_{1}$ with  $\Delta_{0}=0.01\gamma_{1}$ and $\delta=0$ (for upper panel), $\Delta_{0}=\delta=0.01\gamma_{1}$ (for middle panel) and $\Delta_{0}=0.03\gamma_{1}$ and $\delta=0.01\gamma_{1}$ (for bottom panel). (Left panel): The energy dependence of the transmission probability for barrier widths  $d = 10$ nm (blue), $d = 25$ nm (red), and $d = 100 $ nm (green). (Right panel): The width dependence of the transmission probability for incident energies  $E =\frac{1}{5}V_{0}$ (blue), $E =\frac{2}{5}V_{0}$ (red) and $E =\frac{8}{5}V_{0}$ (green).} \label{fig0203}  
\end{figure}

%%%%%%%%%%%%%%%%%%%%%%%%%%%%%%%%% 
\label{NUMERICAL RESULTS AND DISCUSSION}
In this section, we numerically analyze and discuss our main results. First, we evaluate the transmission probability in the two-band model at normal incidence (i.e. $k_{y}=0$).  To understand our system more effectively in  
Fig. \ref{fig0203},
we present the effect of  the band gap $\Delta_0$ on the transmission as a function of the incident energy $E$ and the width $d$ of the barrier.
 %for different values of the system as presented in  . 
In the (left panel), we plot the  energy dependence of the transmission probability for a barrier of width  $d = 10$ nm, $d = 25$ nm, and $d = 100 $ nm for biased $\delta=0$ and unbiased system $\delta\neq0$ with band gap $\Delta_0$.  For  $\Delta_0\neq 0$,  we observe appearance of resonances  in the transmission probability for the energy range $E<V_{0}-\delta'$, $\delta'=\delta+\Delta_0$,
%see. Fig. \ref{fig0203} (left panel), 
which can be attributed %and that is attributable 
to the finite size of the AB-BLG as well as the presence of charge carriers with different chirality. % in the energy range of energies considered.  
These phenomena are known as Fabry-Pérot resonances \cite{Snyman}. For the energy range $V_{0}-\delta'<E<V_{0}+\delta'$, there is a bowl (window) of zero transmission for $d=100$ nm in contrary for $d=10$ nm and $d=25$ nm the transmission is not zero. However, for $E>V_{0}+\delta'$, the transmission still looks like Ben {\it et al.} results \cite{Ben}. Note that the transmission of width $d=100$ nm, shows anti-Klien tunneling, which is a direct consequence of the pseudospin conservation in the system. In the (right panel), we plot the width dependence of the transmission probability %for barrier of width $d = 10$ nm, $d = 25$ nm, and $d = 100 $ nm for 
for the incident energies  $E =\frac{1}{5}V_{0}$, $E =\frac{2}{5}V_{0}$ and $E =\frac{8}{5}V_{0}$.
It is clearly seen that for $E=\frac{1}{5} V_{0}$ and $E=\frac{2}{5} V_{0}$ with $\delta_0=0$, $\Delta_0=0.01\gamma_{1}$, resonance peaks show up (see upper panel), which are absent for the case  $\Delta_0=0$ \cite{Ben}. In the middle and bottom panel, by taking into account the effect of a finite bias $\delta=0.01\gamma_{1}$, we observe a decrease of resonance in the transmission probability,  and  more precisely when  $\Delta_0$ is greater than $\delta$. % See Fig. \ref{fig0203} (right panel).

\begin{figure}[ht]
        \begin{center}
        \includegraphics[width=0.742\linewidth]
        {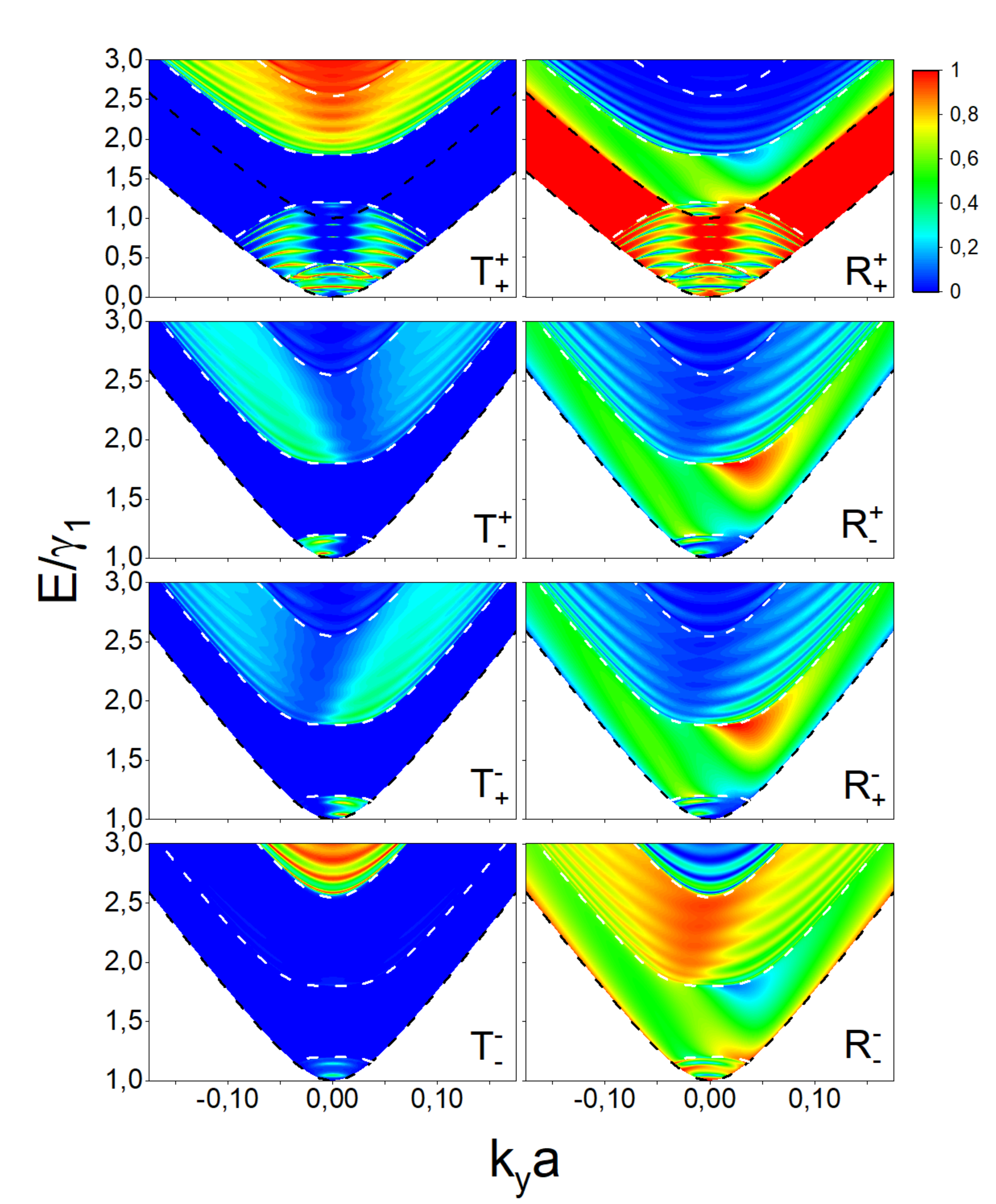}
        \end{center}
        \caption{(Color online) Density plot of transmission and reflection probabilities as a function of  the incident energy $E$ and  transverse wave vector $k_{y}$, through a potential barrier of height $V_{0} =1.5\gamma_{1}$ and width $d=25$ nm and band gap $\Delta_{0}=0.3\gamma_{1}$ with $\delta=0$. The dashed white and black lines represent the band inside and outside the barrier, respectively.}
        \label{fig0204}
\end{figure}

To investigate the effect of band gap, for energy greater than the  interlayer hopping parameter, $E>\gamma_{1}$, in
Fig. \ref{fig0204}
we show the transmission and reflection channels as a function of the  incident energy $E$ and  transverse wave vector $k_{y}$ for  potential height  $V_{0}=\frac{3}{2}\gamma_{1}$ and  width $d=25$ nm.
 The superimposed dashed curves indicate different propagating modes inside and outside the barriers. %Ben  {\it et al.}\cite{Ben} 
For ungapped and unbiased AB-BLG (pristine AB-BLG), Ben  {\it et al.} \cite{Ben} showed that all channels are symmetric with respect to normal incidence, $k_y=0$, i.e. 
% In other words, all channels of the transmission and reflection are equivalent meaning that 
$T^{+}_{-}=T^{-}_{+} $ and $R^{+}_{-}=R^{-}_{+}$.
{This is due to the valley equivalence, namely the transmission probabilities of electrons moving in the opposite direction (scattering from $k^{+}$ to $k^{-}$ in the vicinity of the first valley, and scattering from $k^{-}$ to $k^{+}$ in the vicinity of the second valley) are the same}. 
%Also, it is noticed that  the cloak effect occurs at   $k_y=0$ for pristine AB-BLG. 
Now as for our case by introducing a gap $ \Delta_{0}=0.3 \gamma_{1}$,  with a null inter-layer bias, $\delta=0$, 
	%the energy bands in Eq. (\ref{energy}) reduced to $\varepsilon^{s}_\pm=s\sqrt{k^{2}+\Delta_{0}^{2}+\frac{1}{2}\pm\sqrt{k^{2}+\frac{1}{4}}}$, and clearly show that the nature behavior of the gap is different from that the biased AB-BLG. Where is she showing up as ``Bows" as seen in Fig. \ref{fig0204}. Wheres for $\Delta_0=0$ and $\delta=0.3\gamma_{1} $, i.e $\varepsilon^{s}_\pm=s\sqrt{k^{2}+\delta^{2}+\frac{1}{2}\pm\sqrt{k^{2}(1+4\delta^{2})+\frac{1}{4}}}$, she showing up as ``Mexican hats"\cite{Ben}. Consequently, the band gap $\Delta_0$ should affect the transmission with another way. In fact, 
	we observe that the transmissions  are completely suppressed in the energy range   $V_{0}-\Delta_{0}<E<V_{0}+\Delta_{0}$ due to the absence of traveling modes. In $T^{+}_{+}$ channel and for energies smaller than $V_{0}-\gamma_{1}$, we find that the resonances are decreased and  Klein tunneling get less incandescent than that seen in \cite{Ben}. 
 % and due to the absence of traveling modes,% the transmission channels are completely suppressed in the energy range   $V_{0}-\Delta_{0}<E<V_{0}+\Delta_{0}$.
 We notice that there is asymmetric in the transmission  channels with respect to normal incidence,  $T^{+}_{-}(k_{y})=T^{-}_{+}(-k_{y})$, but reflection channels still showing symmetric behavior,  $R^{+}_{-}(k_{y})=R^{-}_{+}(k_{y})$, because the incident electrons  back again in an electron state \cite{Ben}. %, as seen in the Fig. \ref{fig0204}. 
This is not the  case for gapped AA-BLG, whereas  $T^{+}_{-}$ and $T^{-}_{+}$ channels preserve the momentum symmetry \cite{Bahlouli}.  
	In addition, there is a significant distinction for all reflection channels, $R^{s}_{\pm}$, between gapped AB-BLG and biased AB-BLG. Indeed,
	in our case we observe that the scales of $R^{s}_{\pm}$ get reduced inside the barrier. It is remarkably seen that our transmission channels, $T^{s}_{\pm}$, showed some bowels in the energy spectrum instead  of ``Mexican hats"  as have been see  in \cite{Ben}. This show that $\Delta_0$
	can be used to control the transmission behavior in AB-BLG.
	%
	%We observe that the transmission graduations inside the barrier no longer exist in the case of gapped AB-BLG and in contrary to biased AB-BLG
	%there is no . }
   
\begin{figure}[ht]
        \begin{center}
        \includegraphics[width=.742\linewidth]
        {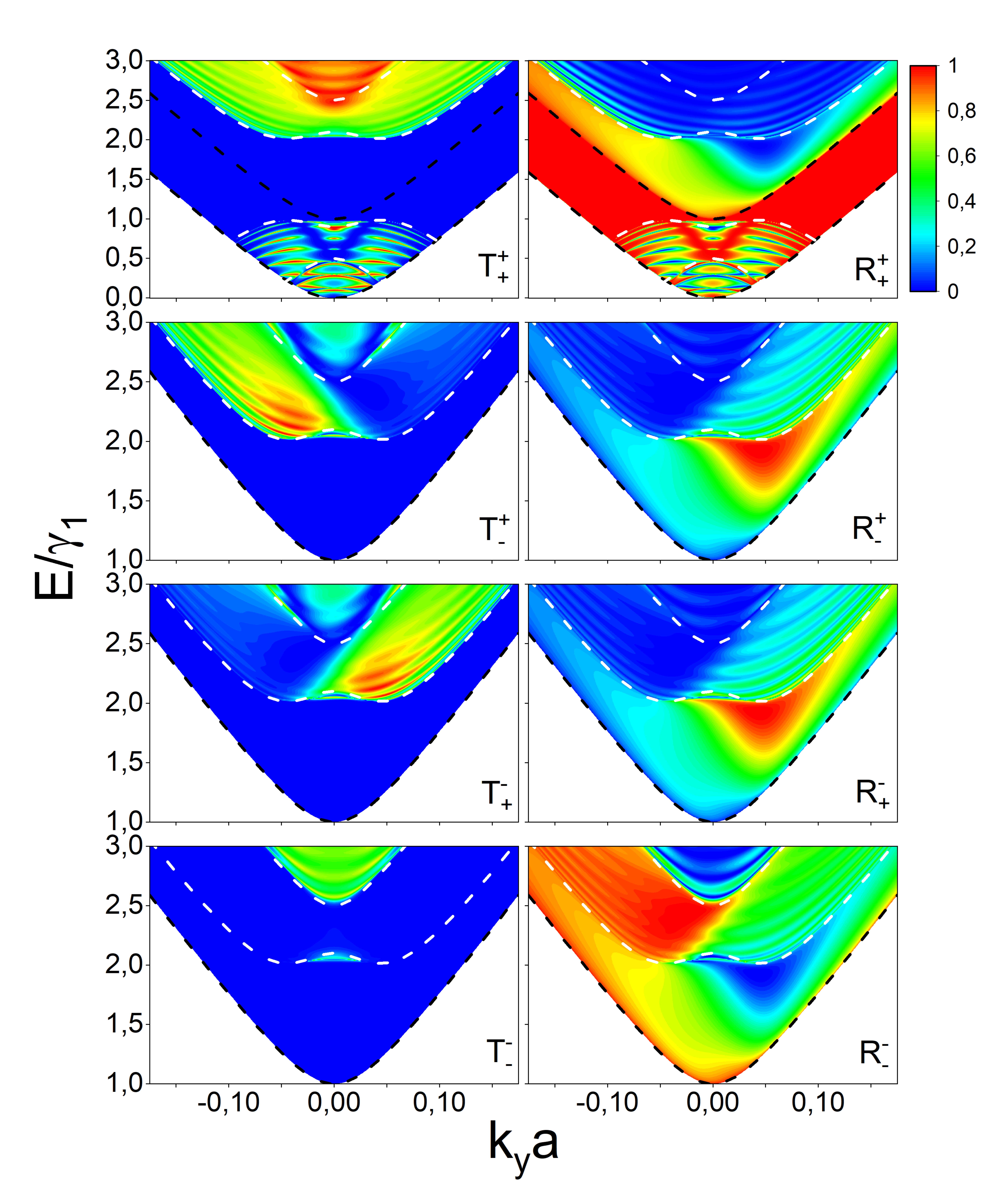}
        \end{center}
        \caption{(Color online) The same as in Fig. \ref{fig0204}, but now for the band gap $\Delta_{0}=0.3\gamma_{1}$ with $\delta=0.3\gamma_{1}$. The dashed white and black lines represent the band inside and outside the barrier, respectively.}
        \label{fig0205}
\end{figure}

In Fig. \ref{fig0205} we show the density plot of the transmission and reflection channels, for biased and gapped systems, $\delta={0.3}\gamma_{1}$, $\Delta_{0}=0.3 \gamma_{1}$. The transmission is  completely suppressed in the energy range  $V_{0}-\delta'<E<V_{0}+\delta'$,  $\delta'=\Delta_{0}+\delta$ %due to the absence of propagating modes
. We notice that the symmetric inter-layer sublattice equivalence is also broken in this case as seen in Fig. \ref{fig0204}. We recall  that such symmetry broken can be achieved by taking  either $\delta\neq0 $ or $ \Delta_{0}\neq0 $, which means that there is violation of  invariance under the exchange $k_{y}\longrightarrow-k_{y}$ %and then leaves the Hamiltonian (\ref{eq2})  non invariant
as noted in \cite{Ben,Nilsson} for AB-BLG, in contrast to the AA-BLG \cite{Bahlouli1}. Therefore, the  transmission and reflection probabilities are not symmetric with respect to normal incidence as seen in Fig. \ref{fig0205}.

\begin{figure}[ht]
        \begin{center}
       \includegraphics[width=.742\linewidth]
       {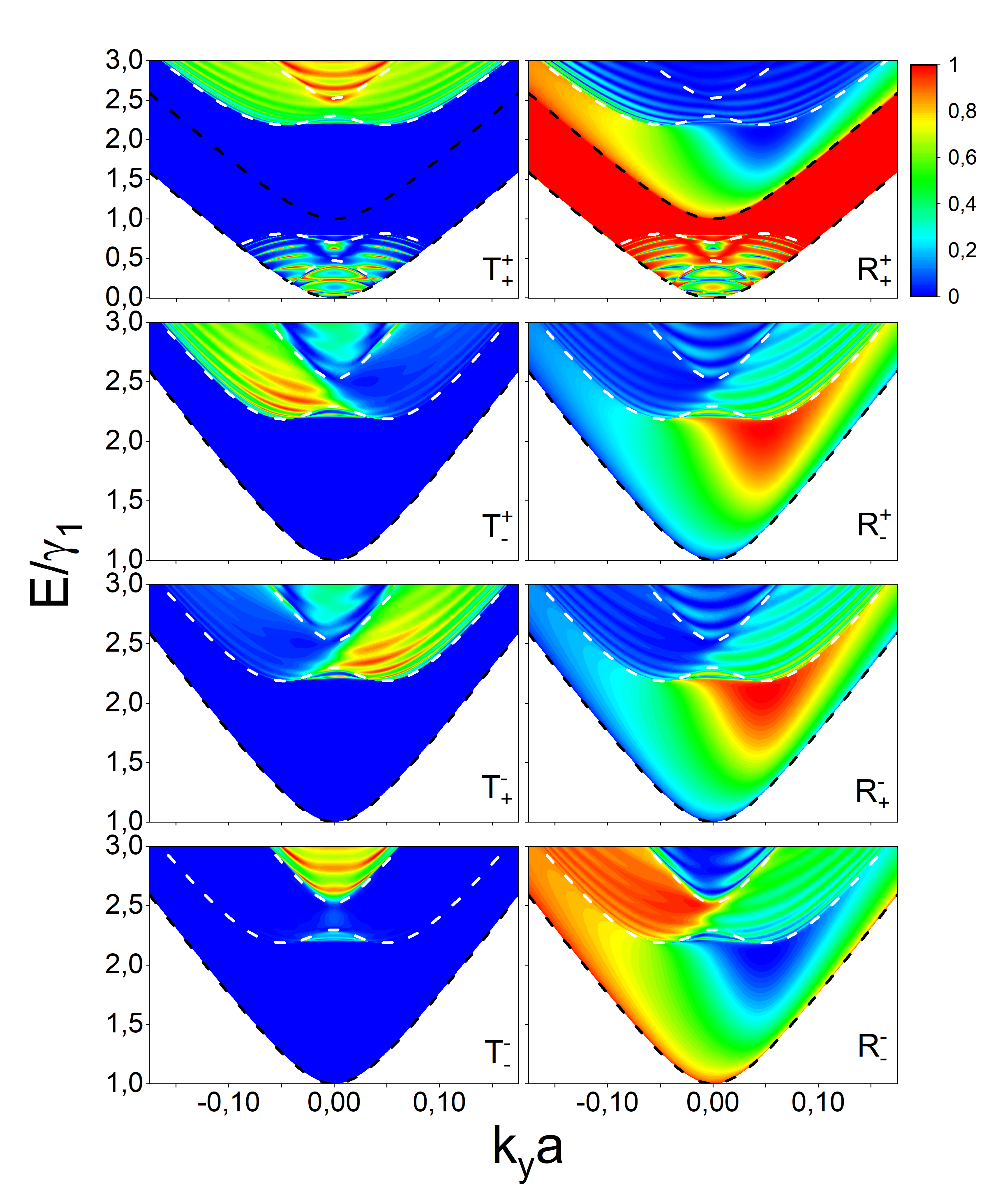}
       \end{center}
       \caption{(Color online) The same as in Fig. \ref{fig0204}, but now for the band gap $\Delta_{0}=0.5\gamma_{1}$ with $\delta=0.3\gamma_{1}$. The dashed white and black lines represent the band inside and outside the barrier, respectively.}
       \label{fig0206}
\end{figure}

Fig. \ref{fig0206} presents the same plot  as in Fig. \ref{fig0205} except that we choose a band gap  $\Delta_{0}=0.5\gamma_{1}$ greater than inter-layer bias $\delta=0.3\gamma_{1}$. In this situation, we notice a significant difference in the transmission and reflection channels.
Indeed, we observe that  Klein tunneling becomes less than that see  for  the case $\Delta_0=\delta=0.3\gamma_{1}$ in Fig. \ref{fig0205}. In addition, it is clearly seen  that  some resonances disappear for %{\rr à supprimer the range of energies}
 the energy range $E<V_{0}-\delta'$. Moreover, 
 we find that the energy bands are pushed and showed some behaviors look like  ``Mexican hats", which are  more clear than  those see in Fig. \ref{fig0205}. These results are similar to those obtained in \cite{Bahlouli2}, by analyzing 
 the transmission probabilities for a system composed of two single layer-AB bilayer-two single layer (2SL-AB-2SL) of graphene subjected to strong gate potential. In summary, we observe that
  all transmissions for $\delta\neq0$ and $\Delta_{0}\neq0$ are weak compared to the biased AB-BLG \cite{Ben}, or gapped AB-BLG (Fig. \ref{fig0204}) cases.

\begin{figure}[tb]
\begin{center}
\includegraphics[width=9cm]{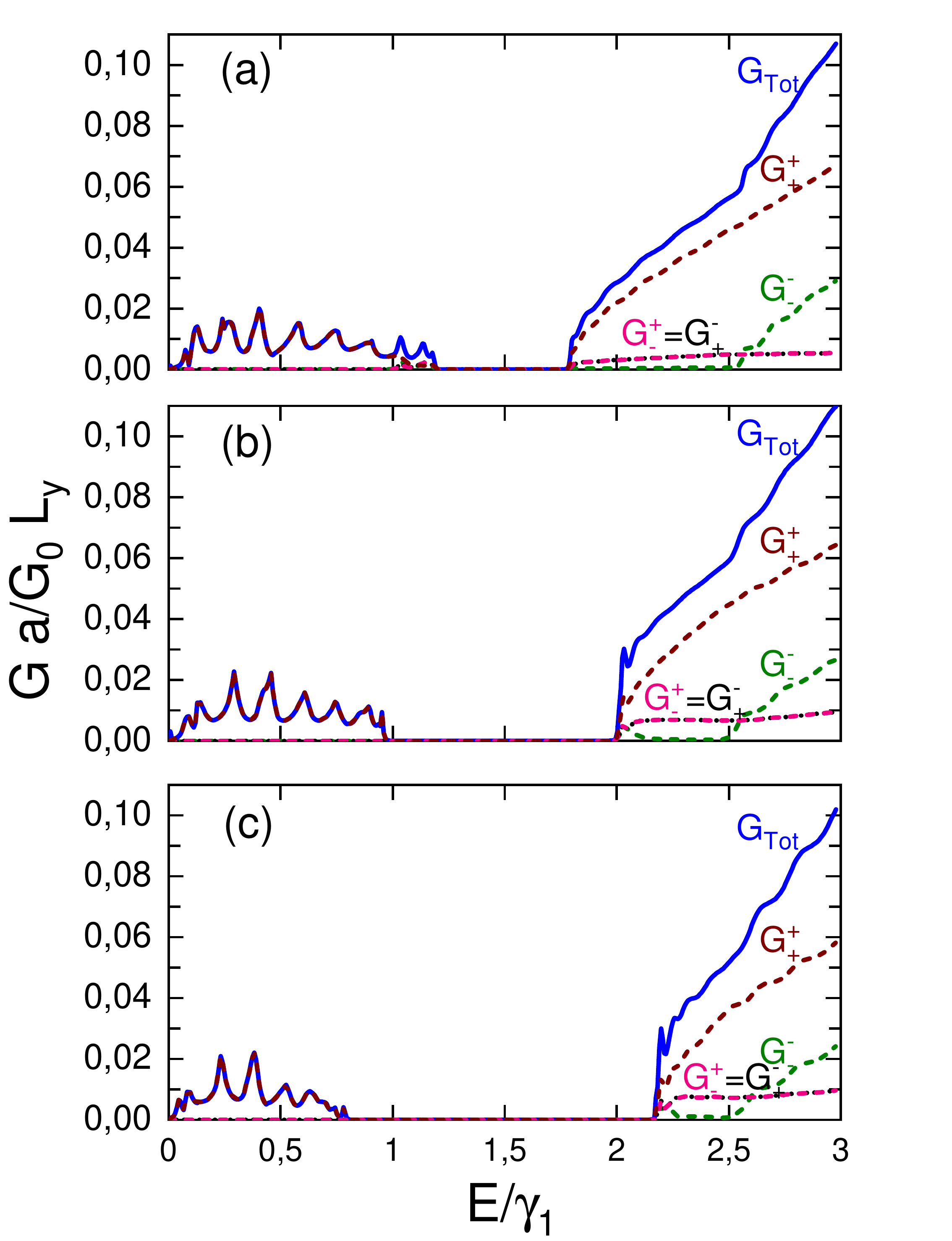}
\end{center}
 \caption{(Color online): Conductance as a function of the incident energy  for biased and gapped AB-BLG 
 with potential height $V_0=1.5 \ \gamma_1$ and width $d=25$ nm. 
 (a):  $\Delta_{0}=0.3\gamma_{1}$, 
 $\delta=0$. (b): $\Delta_{0}=0.3\gamma_{1}$, $\delta=0.3\gamma_{1}$, 
 (c): $\Delta_{0}=0.5\gamma_{1}$, 
 $\delta=0.3\gamma_{1}$. 
 The solid curves correspond to the total conductance  and the dashed curves correspond %\usepackage[utf8]{inputenc}
 to different contributions of the four transmission channels.}\label{fig6} 
\end{figure}

In Figs. \ref{fig6} we plot the energy dependence of the corresponding  conductance for  different values of the band gap and an inter-layer bias $\delta=0.3\gamma_1$. The band gap $\Delta_{0}=0.3\gamma_{1}$ contributed by opening a gap in the energy spectrum of  AB-BLG at $ V_{0} \pm \Delta_{0}$, and this of course reflected on the conductance as shown in Fig. \ref{fig6}(a). The resonances that are clear in the transmission probability show up as peaks, and the total conductance $G_{\text{Tot}}$ has a convex form. For low energies we have $G_{\text{Tot}}=G^{+}_{+}$ meaning that the propagation is only via $k^{+}$ mode, while $k^{-}$ mode is cloaked in this regime until $E> V_{0} + \Delta_{0}$. $G^{-}_{-}$ starts conducting by making an appearance as a rapid increase in the total conductance. Furthermore, $G^{+}_{-}=G^{-}_{+}=0$ since $T^{+}_{-}=T^{-}_{+}= 0$ at low energy but at $E=\gamma_{1}$ both modes are coupled and  $G^{+}_{-}$, $G^{+}_{-}$ start conducting that is why $G_{\text{Tot}}\neq G^{+}_{+}$. However the band gap does not break the equivalence in the scattered channels of the conductance such that $G^{-}_{+}=G^{+}_{-}$ still equivalent for all energy ranges  (see Fig. \ref{fig6}(a)), in contrast to the case of the double barriers \cite{Jellal}. By comparing our results with those of the biased AB-BLG \cite{Ben}, we observe that some shoulders of the peaks are removed and the contribution of the transmission channels on the total conductance are not much more pronounced as a result of the gap $\Delta_{0}$ induced by dielectric layers. This confirms that our $\Delta_{0}$ has a significant impact on the transport properties and differs from that  induced by bias in  AB-BLG \cite{Ben}. Instead of contrast, the total  conductance of a gapped AA-BLG is approximately unchanged even though 
the band gap has a significant impact on the intracone transport \cite{Bahlouli}. Now
we involve both of parameters by presenting Figs \ref{fig6}(b) and \ref{fig6}(c) corresponding, respectively, to $\Delta_0=\delta=0.3\gamma_{1}$,  and $\Delta_0=0.5\gamma_{1}, \delta=0.3\gamma_{1}$.  As expected we observe large suppression of the conductance in the  energy range  $V_{0}-\delta'<E<V_{0}+\delta'$, and hence some peaks are removed with a decrease of the total conductance $G_{\text{Tot}}$.

\section{Summary and conclusion}
\label{Summary and conclusion}
We have theoretically investigated the transport properties through rectangular potential barriers of biased AB-BLG gapped by dielectric layers. By solving Dirac equation, the four band energies are obtained to be dependent on the band gap $\Delta_{0}$ together with the inter-layer bias $\delta$. 
Subsequently, using transfer matrix method we have evaluated the corresponding transmission, reflection probabilities, and  conductance. In particular, we have analyzed the transmission probability in the two-band model at normal incidence, (i.e $k_{y}=0$), firstly in the presence of  $\Delta_{0}$ and secondly by taking into account   $\Delta_{0}$ and $\delta$. As a result, we have observed that the presence of $\Delta_{0}$ induces extra resonances appearing in transmission profiles. However by adding $\delta$, we have observed that the transmission decreased more 
and  anti-Klein tunneling in AB-BLG is no longer preserved.

Furthermore, we have obtained a new mode of propagation for energies exceeding the inter-layer coupling $\gamma_{1}$. In this case,  we have showed that  the band gap $\Delta_{0}$ breaks the inter-layer sublattice equivalence with respect to $k_{y}=0$. Such asymmetry is apparent in the scattered transmission where it depends on the incident mode. The corresponding conductance does not incorporate this asymmetric, and the locations of their peaks are  changed in the presence of $\Delta_{0}$ compared to  $\delta$ \cite{Ben}.

\section{Acknowledgments}
The generous support provided by the Saudi Center for Theoretical
Physics (SCTP) is highly appreciated by all authors. A.J. thanks Dr. Michael Vogl for fruitful  discussion.

\end{document}